\documentclass{aa} 
\usepackage{graphicx}
\begin{document}
   \title{A Rigorous Comparison of Different Planet Detection Algorithms}

   \author{B. Tingley
          \inst{1}
          }

   \offprints{B. Tingley}

   \institute{$^{1}$Institut for Fysik og Astronomi (IFA), Aarhus Universitet,
              Ny Munkegade, Bygning 520, 8000 Aarhus C\\
              \email{tingley@ifa.au.dk}
             }

   \date{Received X, 2002; accepted Y, 200Z}

   \abstract{
The idea of finding extrasolar planets (ESPs) through observations
of drops in stellar brightness due to transiting objects has been
around for decades. It has only been in the last ten years, however,
that any serious attempts to find ESPs became practical.
The discovery of a transiting planet around the star HD 209458 (Charbonneau
et al. \cite{charbonneau}) has led to a veritable explosion of research,
because the photometric method is the only way to search a large number
of stars for ESPs simultaneously with current technology. To this point,
however, there has been limited research into the various techniques used
to extract the subtle transit signals from noise, mainly brief summaries
in various papers focused on publishing transit-like signatures in
observations. The scheduled launches over the next few years of satellites
whose primary or secondary science missions will be ESP discovery motivates
a review and a comparative study of the various algorithms used to perform
the transit identification, to determine rigorously and fairly which one 
is the most sensitive under which circumstances, to maximize the results of
past, current, and future observational campaigns.

   }

   \maketitle
%

\section{Introduction}

Struve (\cite{struve}) was the first to postulate that ESPs could be 
found in the event that they transited their parent stars, which would
dim slightly as the ESP occulted some of the light. This
possibility was further explored and developed first by Rosenblatt
(\cite{rosenblatt}) and later by Borucki \& Summers (\cite{borsum}).
Due to technological restraints, the first serious observational
attempts were not performed until Doyle et al. (\cite{doyle0}), who
observed the main-sequence eclipsing binary CM Draconis for the tell-tale
photometric dips that would indicate the presence of a planet. This system
has several qualities that make it an excellent choice for a planet search,
discussed in more detail in Schneider \& Doyle (\cite{schneider}).
First of all, both members are low-mass main sequence dwarfs, which
means that they are physically smaller than a typical field star of
comparable apparent magnitude. This leads, in turn, to deeper transits.
Additionally, the orbital plane of a planet in a binary system is probably
going to be very close to the orbital plane of the stars, increasing the
chance that a planet would cause observable transits. (Barbieri,
Marzari, and Scholl, \cite{barbieri}).
Other projects besides this single, on-going observational campaign
began over the next few years, such as STARE (Brown
\& Charbonneau, \cite{brown}) and VULCAN (Borucki et al. (\cite{borucki}),
which used dedicated small telescopes with large fields of view to observe
thousands of stars simultaneously. This area of research became much more active
after the discovery of the planet around HD 209458 (Charbonneau et al.
\cite{charbonneau}) and
many projects were initiated in the hopes of witnessing an ESP as it
transited its parent star. Among these are several future satellite projects
that list ESP searches as either primary or secondary goals (COROT,
Eddington, Kepler, and MONS to name a few) and HST observations of 47 Tuc
(Gilliland et al. \cite{gilliland}). These observations of 47 Tuc -- like all
of the recent and ongoing projects -- observed a few transit-like events.
All of these events, with one exception, proved to be grazing
eclipses of binary stars after follow-up observations. This one exception
was discovered by
the OGLE group, who were ostensibly searching for optical gravitational
lenses towards the LMC. They observed 59 transit-like signatures in
foreground Milky Way stars, most of which repeated (Udalski et al.
\cite{udalski2}). Of these 59, only one so far has proven in the end
to be an ESP (Konacki et al. \cite{konacki}).

Statistics gathered from high-precision spectroscopic searches for ESPs among
F, G, K and M field stars suggest that approximately 1\% of the observed
objects have giant planets with a period of three to six days. If one considers
that around 10\% of all possible orbital orientations 
of short-period giant planets would produce observable transits,
the expected frequency of detections becomes about one planet per
3000 stars (Borucki et al. \cite{borucki}), provided the effects of daytime
and the three transit requirement are included. If this is so, the paucity of
detections from these various searches is slightly puzzling. It is
understandable, perhaps, that ESPs in clusters could be influenced by the
dense environment there, affecting planet formation and retention -- but
not all of the observing campaigns have been towards clusters.
It is also possible that a statistical prediction of a small number
of ESPs could results in no detections. However, the results from the OGLE
survey, which has only one confirmed ESP out of 50,000 stars (Konacki et al.
\cite{konacki}), suggest that the expected frequency of detections
determined by Borucki et al. is probably too optimistic.  

There are, however, other factors that could affect planet detection. For
example, a poorly conceived or implemented differential photometric reduction
could increase S/N, hiding transits that would otherwise be detectable.
Likewise, the algorithms used to pick the transits out of noise can also
have a strong influence on the number of detections. In fact, the work of
the OGLE group reveals the effect that improvements in
detections algorithms can have. In their first report on the transit-like
events from their observing campaign, they identified 46 events with an
in-house algorithm described only as a 'cross-correlation' (Udalski et al.
\cite{udalski1}). Soon afterwards, they published an update (Udalski et al. 
\cite{udalski2}) listing 59 events -- the new ones identified with an
improved algorithm, the BLS technique described by Kov\'{a}cs et al.
(\cite{kovacs}). There are many algorithms, called detectors in
signal processing literature, which can extract signals from data.
Therefore, it is possible that the seeming scarcity of photometric ESPs
could be caused, at least in part, by the detectors used. 
 
With this possibility raised, a review of the different techniques used
for transit identification is overdue, as is a comparative analysis
to test the various techniques and to determine which is best under which
conditions. Considering the current and future projects that plan to search
for ESPs using photometry, this is a rather important
line of research, one which has not been adequately addressed in the literature
to this point, despite a certain amount of debate on the topic. This study,
and future studies along this line, could potentially have a large impact
on photometric ESP searches, since pushing the high-confidence
detection limit down even a fraction could potentially reveal transits that
would otherwise remain unidentified, shrouded in noise.

\subsection{Transit characteristics}
A planetary transit leaves a very specific signature on the light curve
of a star. As the planet passes across the disk of its parents star,
it occults some of the light. This results in a slight drop in the overall
brightness of the system. The details of the
resulting light curve can easily be calculated to fairly high
precision using a basic limb-darkening law, but it is in fact very well
represented by a simple square well. The depth of the well can be as much as
one or two percent or more, defined approximately by the ratio of the area of
the disk of the ESP and the area of the disk of the parent star. For example,
a transit of Jupiter across the Sun would cause a drop in brightness of
about 1\% -- about a hundredth of a magnitude. The width of the well is
defined by the duration of the transit, which is generally somewhere between
3 and 6 hours for short period planets. This value depends not only on orbital
velocity and radius but also on the impact parameter of the transits, i.e.
how close the ESP's path across the disk of its parent star comes to the
center of the star.

Planet search campaigns using the photometric technique
are far more likely to discover short-period giant planets. Shortening the
period produces more transits in a given period of time, as well
as increasing the chance that a random orbital orientation will produce
an observable transit. This geometrical effect goes as the inverse of the
planetary orbital radius, or, using Kepler's third law, the period of the
orbit to the $-\frac{2}{3}$. This means that, for
example, placing Earth in Mercury's orbit would increase the chance of
transit by a factor of 2.5, while moving it to an orbit with a period of 7
days would increase it by a factor of 14. To skew matters even more in the
favor of short-period planets, short-period giant planets will in theory
be inflated by incident radiation (Guillot et al. \cite{guillot}).
This conclusion is supported by the derived
characteristics of both the planet around HD 209468 (Burrows et al.
\cite{burrows}) and OGLE-TR-56 (Konacki et al. \cite{konacki}). This means
that a short-period planet should be larger than an equally-massed planet
further from the parent star and will therefore create deeper transits.


\section{Transit Identification Algorithms}

A transit identification algorithm is a mathematical tool that examines
light curves for the presence of transits. In general, this is done by
generating a test statistic ($T$) for each set of free parameters in some
fashion or another. If this test statistic exceeds a certain value determined
by the desired level of confidence that the event is not a chance occurrence
of noise ($\eta$), then the algorithm decides that there is a transit-like
event in the light curve. 

Detecting a signal in the presence noise is an old and well-studied problem,
with a variety of fundamental techniques (or detectors) contained
in the literature. However, the application of these various detectors
in the special case of transit identification has not been thoroughly
explored.
While there have been many papers on planetary transit searches published
in the past few years, these have primarily been focused on publishing
results. The descriptions of the detectors used therein to isolate
transit events have been condensed, neglecting a full description of the
nuances that arise in this astronomical context. There are only a few
exceptions. Kay (\cite{kay}) states that a matched filter detector is
the best in the case of Gaussian noise (white or colored, if pre-whitened),
but this is not
necessarily going to be the case in all circumstances, especially considering
all of the processing that is required to prepare data for transit searches.

A literature search reveals a wide variety of detectors that can be
used to identify transits in light curves. The first paper that attempts
to deal with the topic of numerical methods for identifying transits is that
of Jenkins et al. (\cite{jenkins0}). It discusses extensively the use
of a matched filter detector in an attempt to identify transits in the
particularly complicated case of the eclipsing binary CM Draconis. In their
studies, Borucki et al. (\cite{borucki}) and Gilliland et al.
(\cite{gilliland}) used a slight variation on this same technique. In a
further attempt to
deal with the intricacies of CM Draconis, Doyle et al. (\cite{doyle})
developed a different detector. It is also labeled a matched filter
approach, but it is fundamentally different from those used by other groups.
To prevent confusion, this detector will hereafter be distinguished as
Deeg's approach. Defa\"{y} et al. (\cite{defay}) describes a detector
based on Bayesian statistics that not only finds transits, but also
determines the
best-fit shape of the transit. Kov\'{a}cs et al. (\cite{kovacs}) devises
a box-fitting algorithm (BLS), which is based on direct least squares fits
of step functions to the data. This is the only article that attempted
to compare existing detectors. Among the detectors used
in their comparison -- which mostly included ones that were admittedly not
designed for the identification of periodic events with short durations --
was the Bayesian detector from Defa\"{y} et al. (\cite{defay}), which did
not fare favorably. Their article did not, however, include the matched
filter approach in the analysis, nor were their comparisons completely
rigorous, as they involved only a single simulated light curve.

\subsection{The matched filter approach}
The matched filter approach is one of the fundamental tools used in signal
processing. According to Kay (\cite{kay}), this is the optimal detector
for a known signal with white Gaussian noise (WGN). Its conceptual basis
is a calculation of the probability that a set of observations
is the result of WGN or the result of an underlying signal $S_{n}$ plus
WGN. The test statistic is formed by taking the ratio of the noise
probability and the signal + noise probability, which is known as the
likelihood ratio. Taking the natural logarithm of this ratio converts the
product of the probabilities
into a summation and cancels out the exponentials from the Gaussian probability
density functions. After shuttling some constants around, the basic
formulation of its test statistic $T$ becomes (see Kay (\cite{kay}) for
a full derivation):
\begin{equation}
T = \sum_{n=1}^{N}\frac{D_{n}S_{n}}{\sigma_{n}^{2}}
\end{equation}
where $D$ are the observed magnitudes, $S$ is the test signal, $\sigma_{n}$
is the S/N for each observation, and $N$ is the number of observations in the
light curve. $S$ represents only a single set of test parameters, so in order
to search for any transit in the observations, a $S$ must be repeatedly
modified so that parameter space can be covered. A test statistic is then
generated for each test $S$, with the highest value of the test statistics
corresponding to the most likely set of test parameters. 

Implementation of the matched filter approach for transit identification
can be significantly more complex than the basic formulation of the
matched filter. This becomes readily apparent in the case of CM Draconis.
First of all, the binary nature of the system not only affects the light
curve with its regular eclipses, but also severely complicates the pattern of
transits of any planet which might be present (Jenkins et al. 
\cite{jenkins0}). Secondly,
the data are taken at several different telescopes, each with its own CCD
camera -- or even just two-star photometers in one case -- meaning that
the quality of the data will vary systematically (Deeg et al.
\cite{deeg}), further muddling time series. Thirdly, all of the good comparison
stars in the field were significantly bluer than CM Draconis, which
resulted in residual nightly extinction variations
(Doyle et al. \cite{doyle}). Clearly, a simple matched filter approach
is not sufficient for this circumstance. Jenkins et al. (\cite{jenkins0}),
cleaned the data to a high degree by subtracting model eclipses.
However, this still left the problem with the residual nightly
extinction variation, which led the same group to develop another detector.

\subsection{Deeg's approach}

Deeg's approach is also based on the idea of comparing the data with
a series of test signals spanning parameter space. It was specifically
designed for the complex circumstances found in a data set of several years
of observations of CM Draconis. It differs fundamentally from the matched
filter approach described above in that it includes a time-based weighting.

In order to generate their test statistic, Doyle et al. (\cite{doyle})
generated a test signal with transits included (the with-planet test signal).
They subtracted this with-planet test signal from the light curve for each
individual night of observations and fit a parabola to what remained. Then
they fit a parabola to each individual night of observations, which is intended
to model the residual nightly extinction variation. These fits are then
compared to the original data and the residuals determined. From these
residuals, the test statistic is calculated:
\begin{equation}
T = \sum_{n=1}^{N}\kappa_{n}
\end{equation}
where
\[\kappa_{n} = \left\{ \begin{array}{ll}
0 & \mbox{if $t_{n+1}-t_{n}>10$ min} \\
(r^{e}_{n}-r^{p}_{n})/\Delta t\sigma_{D} & \mbox{otherwise}
\end{array}
\right. \]
with
\[\Delta t = (t_{n+1}-t_{n}), \]
\[r^{e} = |D - f^{e}|\]
and
\[r^{p} = |D - S - f^{p}|,\]
where $f^{e}$ is the best-fit parabola in the no-planet case, $f^{p}$
is the best-fit parabola in the with-planet case, $\sigma_{D}$ is the
noise level for the night analyzed, and $t_{n}$ is the time of the
observation. The
inclusion of time in this detector is unique, effectively weighting
each element of the test statistic by the time between consecutive
observations. If this $\delta t$ is too long, however, the points
are neglected, as these regions of the light curve correspond to ``holes''
in the data. According to Doyle et al. (\cite{doyle}), this time-based
weighting was necessary to account for the difference in time increments
of the observations from different telescopes.

\subsection{Bayesian approach}

The Bayesian approach described by Defa\"{y} et al. (\cite{defay}) is
based on a different statistical philosophy than the previously mentioned
techniques. In a nutshell, it estimates an unknown
parameter through the maximization of a likelihood function, invoking
as much prior information as possible in order to improve the estimation.

The primary parameter that Defa\"{y} et al. fit is the period of the signal
to be detected. They assume that the noise was WGN and then represented the
signal as an unknown Fourier series. By finding the most likely period of
the signal, the coefficients of the Fourier series can then be determined,
fitting the actual shape of the transit. The likelihood function that they
derive, to be maximized for frequency corresponding to the period of the
signal, is
\begin{equation}
\log L(\omega) \propto \frac{-N}{2\sigma^{2}}
  \sum_{k=1}^{m}\left(\frac{\alpha_{k}^{2}}{N^{2}} +\frac{\beta_{k}^{2}}{N^{2}}\right) 
\end{equation}
where
\begin{equation}
\alpha_{k} = \sum_{n=1}^{N}s_{n}\cos(\omega kt_{n}),\qquad k = 1...m
\end{equation}
\begin{equation}
\beta_{k} = \sum_{n=1}^{N}s_{n}\sin(\omega kt_{n}),\qquad k = 1...m
\end{equation}
$s_{n}$ is the $n^{th}$ point in the light curve, $t_{n}$ is its
corresponding time, and $\omega$ is the frequency. In
theory, $m \rightarrow \infty$ is necessary to fit the transit shape
properly. However, the authors truncate the summations of $\alpha_{k}$ and
$\beta_{k}$ at $m = 7$ to reduce the considerable computational load of
this approach, claiming that this can be safely done without loss of
precision. After the equation has been maximized and the most likely
frequency determined, the shape of the transit can be reconstructed:
\begin{equation}
f(t) = \sum_{k=1}^{m}\left(\frac{2\alpha _{k}}{N}\cos(\omega kt)+\frac{2\beta _{k}}{N}\sin(\omega kt)\right).
\end{equation}
It is important to remember that this implementation
of the detector works only for data without ``holes'', although it could
be modified for the case of unevenly sampled data.

\subsection{The Box-fitting technique}

The box-fitting method described by Kov\'{a}cs et al. (\cite{kovacs}) is
essentially a $\chi^{2}$ fit of a square-well transit model to the
observations. Through minimization, they are able to remove the depth of
the transit as a free parameter, reducing the computational load.
The expression to be minimized for a given $n_{1}$ and $n_{2}$ is
\begin{eqnarray}
D = \sum_{n=1}^{n_{1}-1}w_{n}(D_{n}-H)^{2}+\sum_{n=n_{1}}^{n_{2}}w_{n}(D_{n}-L)^{2} \nonumber \\
+\sum_{n=n_{2}+1}^{N}w_{n}(D_{n}-H)^{2},
\end{eqnarray}
where $w_{n} = \sigma_{n}^{-2}[\sum_{m=0}^{N-1}\sigma_{m}^{-2}]^{-1}$ is the
normalized weight of each data point, $L$ is the signal during transit
(between $n=n_{1}$ and $n=n_{2}$), and $H$ is the signal outside the transits.
The authors then make the assumption that the average of the $D$ is
zero. This allows them to reduce the number of parameters from five ($P_{0}$,
the period of the repeating transit, $q$, the fractional transit length, $L$,
$H$, and $t_{0}$, the epoch of the transit) to four with the substitution of
$H = -Lq/(1-q)$. This is accurate provided that $q$ is small. This
substitution allows $L$
to be parameterized as $\frac{s}{r}$ and $H$ as $-\frac{s}{1-r}$, where
$s=\sum_{n=n_{1}}^{n_{2}}w_{n}D_{n}$ and
$r=\sum_{n=n_{1}}^{n_{2}}w_{n}$. The expression to be minimized then becomes
\begin{equation}
D=\sum_{n=1}^{N}w_{n}D_{n}^{2} - \frac{s^{2}}{r(1-r)}.
\end{equation}
This particular form of the equation is useful.
Not only does the first term not depend on n  and n  , allowing it to be
dropped, but neither of the terms include the depth of the transit,
creating a test statistic that is now independent of this free parameter:
\begin{equation}
T = {\rm max}\left\{\left[\frac{s^{2}({n_1},{n_2})}{r({n_1},{n_2})-[1-r({n_1},{n_2})]}\right]\right\}^{\frac{1}{2}}.
\end{equation}
   \begin{figure*}
   \centering
   \includegraphics{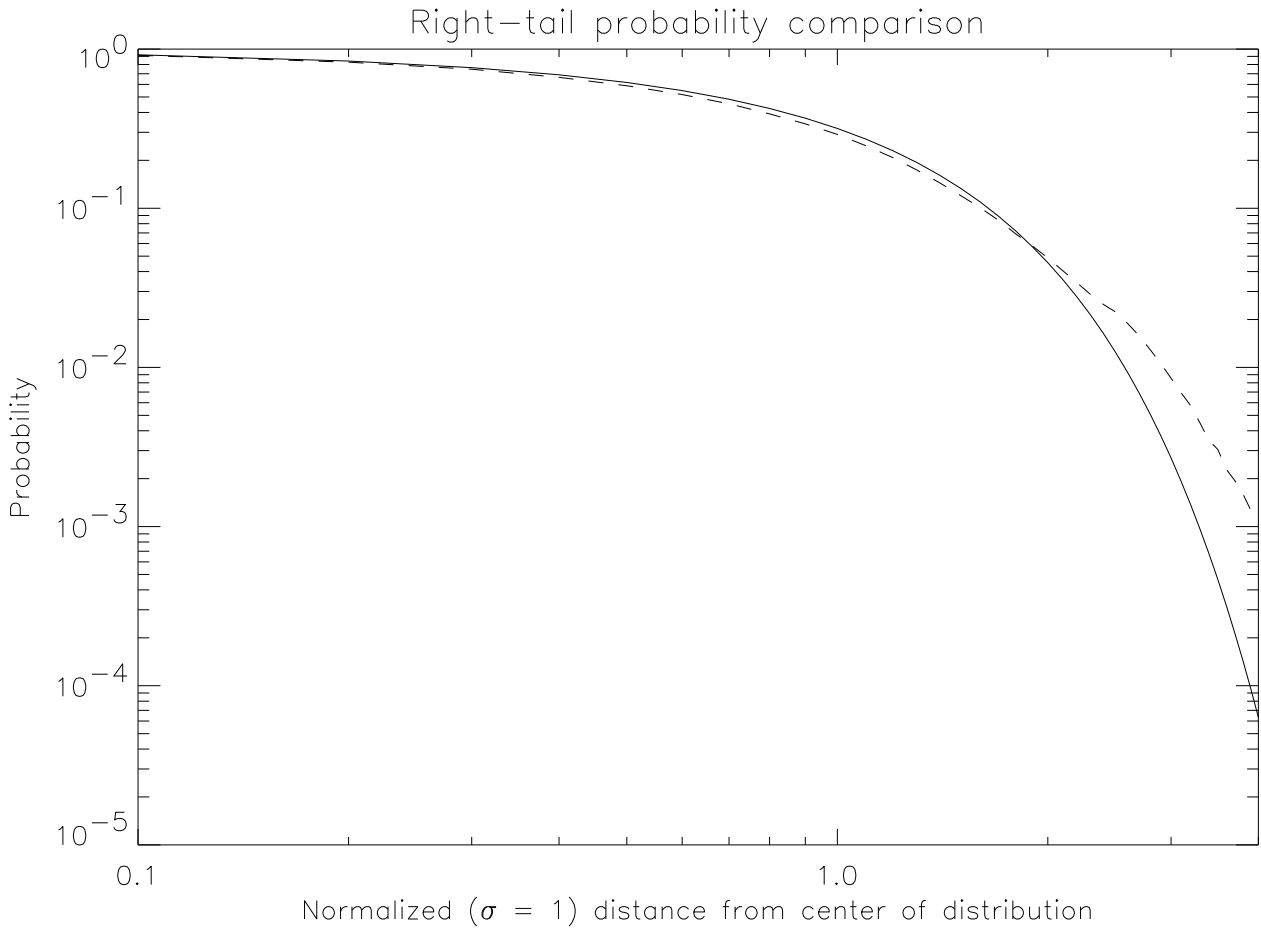}
   \caption{This figure shows the comparison between the distribution of
the observations used to generate the virtual light curves (dashed line)
and a Gaussian distribution (solid line). Notice
the excess where the normalized distance from the center of the
distribution is greater than $1.1$. Despite the fact that it looks large, it is
actually due to just $10$ out of the $10^{4}$ observations.}
    \end{figure*}

\subsection{Correlation}

The correlation is a very basic statistical tool used to measure how well
two samples resemble each other. It is not included here as a serious transit
identification method, but instead primarily as a point of reference.
Since it is not designed for transit identification,
it can be used as a benchmark to measure the effectiveness of the
statistical philosophies behind the various detectors. The textbook
definition of the coefficient of linear correlation is
\begin{equation}
T = \frac{\sum_{n=0}^{N-1}(D_{n}-\bar{D})(S_{n}-\bar{S})}
         {[\sum_{n=0}^{N-1}(D_{n}-\bar{D})^{2}\sum_{n=0}^{N-1}
          (S_{n}-\bar{S})^{2}]^{\frac{1}{2}}},
\end{equation}
where $\bar{D}$ and $\bar{S}$ are the averages of the light curve and
the expected signal, respectively.

\subsection{Aigrain's Approach}

Aigrain \& Favata (\cite{aigrain}) take the same Bayesian philosophies used
by Defa\"{y} et al. (\cite{defay}) and extend them to assume square well
transits. Ultimately, the mathematical basis proves to be very similar to
the matched filter, different primarily in the way that parameter space is
explored -- the matched filter makes cuts across parameter space, while the
Bayesian ultimately integrates over parameter space. While this is
interesting in and of itself, the limited number of parameters explored in
the simulations performed in this analysis would not be sufficient to examine
the effect of this difference. For this reason, no attempt to include this
approach in this paper is made.

\section{Comparison Method}

In order to compare the various transit identification techniques,
virtual light curves should be generated in as realistic a manner as
possible and the most influential parameters must be identified. In
addition, a fair method to compare the results of the different techniques
must be applied in order to form a definitive conclusion.

Several parameters are needed to characterize the signal caused by
a transiting ESP and more to characterize the observations of one. The
transit itself can be defined by a few physical parameters: depth, width,
time of the first observed transit and period. Including the observations
requires more: the S/N of the
observations, the rate of the observations and the time coverage of the
observations. This proves to be an unwieldy number of parameters to simulate,
so the period is fixed at 3 days and the width at 3 hours. Additionally,
the time of the first transit is also held constant, as it proved not to
make any difference. The depth of the
transit and the S/N are combined into one parameter, the normalized depth
($d$). The period and the time coverage are also combined
into one parameter, the number of transits observed ($N_{{\rm tr}}$).
This leaves only the rate of observations ($r_{{\rm obs}}$). By varying these
last three parameters, parameter space is covered to a sufficient extent for
an evaluation of detector performance. 

The standard technique to construct synthetic light curves is to use a random
number generator that produces WGN for the virtual observations. This
is an assumption that does not necessarily represent the properties of
the noise in real data. In order to avoid this, the virtual light curves are
created by drawing individual observations (as opposed to short sequences of
consecutive observations) randomly from a set of more than $10^{4}$
real observations. This set is derived from observations of
NGC 6791 by Bruntt et al. (in preparation), consisting of data from 17 stars,
each non-variable, from a small region of the main sequence, and with variances
between $4.95\times10^{-5}$ and $6.10\times10^{-5}$. The light curve for
each star was $\sigma$-clipped, first to remove extreme outliers and
again to remove more intermediate outliers that ultimately reduced the
furthest outlier to slightly more than $4\sigma$. After this, each light
curve was individually normalized to create the combined set.

It is important to determine how closely this set of observations matches
WGN in order to understand how globally applicable the results from this
article can be. One way of doing this is to compare its right-tail
probabilities with that of the best-fit Gaussian. As the values in the
set are already normalized, a direct comparison to the complementary
cumulative distribution function ($\frac{1}{2}-\frac{1}{2}{\rm erf}
\left(\frac{x}{\sqrt{2}}\right)$, where ${\rm erf}(x)$ is the error
function) is sufficient. As can be seen in Figure 1, there is an excess
far out in the wings of the distribution, which can be explained by
approximately 10 outlying observations. This is a very small percentage of
the total, meaning that the distribution strongly resembles the Gaussian
without being identical, which is typical of real observations.

While the study of individual light curves might be instructive on a case
by case basis, a Monte Carlo simulation is truly needed to compare the
different detectors reliably. For each set of transit
parameters, $10^{4}$ light curves are generated. The temporal spacing
of the simulated observations is allowed to vary up to 10\% of the average
spacing, with no large gaps representing daytime included - although in
principle this would only affect the current implementation of the Bayesian
approach.

In order to compare the transit identification algorithms, each of the
randomly generated light curves are copied, then transits are added
to the copy. Each of the detectors
are then applied to both copies of the light curve, scanning phase
space for transits identical to those added on, generating
a test statistic for each phase. The peak of these test statistics
corresponds to the most likely phase of the transits. In the case of the
with-transit light curves, this is usually the inserted
transits, depending on the response of the detector to the transits
included. For the light curves without transits included,
the peak is the strongest false alarm -- a false alarm being defined as
a chance occurrence of noise that to a greater or lesser extent resembles
the sought-after signal. These peaks can
then be gathered and the resulting distributions analyzed to determine how
well the algorithms identify transits. It should be noted that there are
instances when a given detector does not find transits even though they
are present. These instances will be increasingly more common as the
transit signals become weaker. This means that the with-transit distributions
increasingly resemble the without as the transit signals weaken.

There are two standard ways of comparing the performance of detectors. One
is through an examination of Receiver Operating Curves
(ROCs), which plot
the chance of detection against the chance of false alarm. The better the
detector is, the higher it lies in this type of diagram. The worst case
scenario - simple guessing - is represented by a straight line with
unity slope. These diagrams are
easily made with a simple calculation from the with-transits and
without-transit distributions for individual detectors and sets of
transit parameters. Another way to compare detectors is to calculate the 
probability of detection at a given probability of false alarm. This is
done simply by ranking the without-transit distributions, determining
the detection thresholds corresponding to the desired false alarm
probabilities (FAPs), and calculating
the portions of the with-transit distributions that exceeds these values.

\begin{table*}
 \centering
 \caption[]{Probabilities of detection for a false alarm rate of 1\% for
various transit parameters and detectors. $N_{tr}$ is the number of transit,
$r_{{\rm obs}}$ is the rate of observations in observations per hour and
$d$ is the depth of the transit in normalized magnitude -- effectively
the S/N of the observations.}
 \begin{tabular}{ccc|ccccc}
  \hline
\multicolumn{3}{c|}{Transit parameters} & \multicolumn{5}{c}{Probability of Detection} \\
$N_{tr}$ & $r_{{\rm obs}}$ & $d$ & BLS & Bayesian & Correlation & Matched Filter & Deeg's \\
  \hline
 3 &  5 & 0.10 & 0.0117 & 0.0105 & 0.0110 & 0.0131 & 0.0122 \\
 3 &  5 & 0.25 & 0.0186 & 0.0155 & 0.0341 & 0.0434 & 0.0380 \\
 3 &  5 & 0.50 & 0.1393 & 0.0500 & 0.3920 & 0.4590 & 0.3617 \\
 3 & 10 & 0.10 & 0.0109 & 0.0111 & 0.0160 & 0.0175 & 0.0156 \\
 3 & 10 & 0.25 & 0.0359 & 0.0244 & 0.1105 & 0.1319 & 0.0887 \\
 3 & 10 & 0.50 & 0.5607 & 0.1445 & 0.8454 & 0.8867 & 0.8014 \\
 6 &  5 & 0.10 & 0.0110 & 0.0111 & 0.0158 & 0.0181 & 0.0152 \\
 6 &  5 & 0.25 & 0.0224 & 0.0179 & 0.1148 & 0.1451 & 0.1084 \\
 6 &  5 & 0.50 & 0.4846 & 0.1136 & 0.8574 & 0.8996 & 0.8193 \\
 6 & 10 & 0.10 & 0.0123 & 0.0123 & 0.0251 & 0.0294 & 0.0225 \\
 6 & 10 & 0.25 & 0.0974 & 0.0339 & 0.3882 & 0.4406 & 0.3052 \\
 6 & 10 & 0.50 & 0.9528 & 0.4460 & 0.9986 & 0.9993 & 0.9951 \\
10 &  5 & 0.10 & 0.0101 & 0.0113 & 0.0202 & 0.0232 & 0.0184 \\
10 &  5 & 0.25 & 0.0434 & 0.0216 & 0.3006 & 0.3392 & 0.2311 \\
10 &  5 & 0.50 & 0.8064 & 0.3017 & 0.9911 & 0.9951 & 0.9863 \\
10 & 10 & 0.10 & 0.0122 & 0.0118 & 0.0391 & 0.0478 & 0.0342 \\
10 & 10 & 0.25 & 0.2234 & 0.0595 & 0.7325 & 0.7898 & 0.6319 \\
10 & 10 & 0.50 & 0.9983 & 0.8371 & 1.0000 & 1.0000 & 1.0000 \\
 \hline
 \end{tabular}
\end{table*}

   \begin{figure*}
   \centering
   \includegraphics{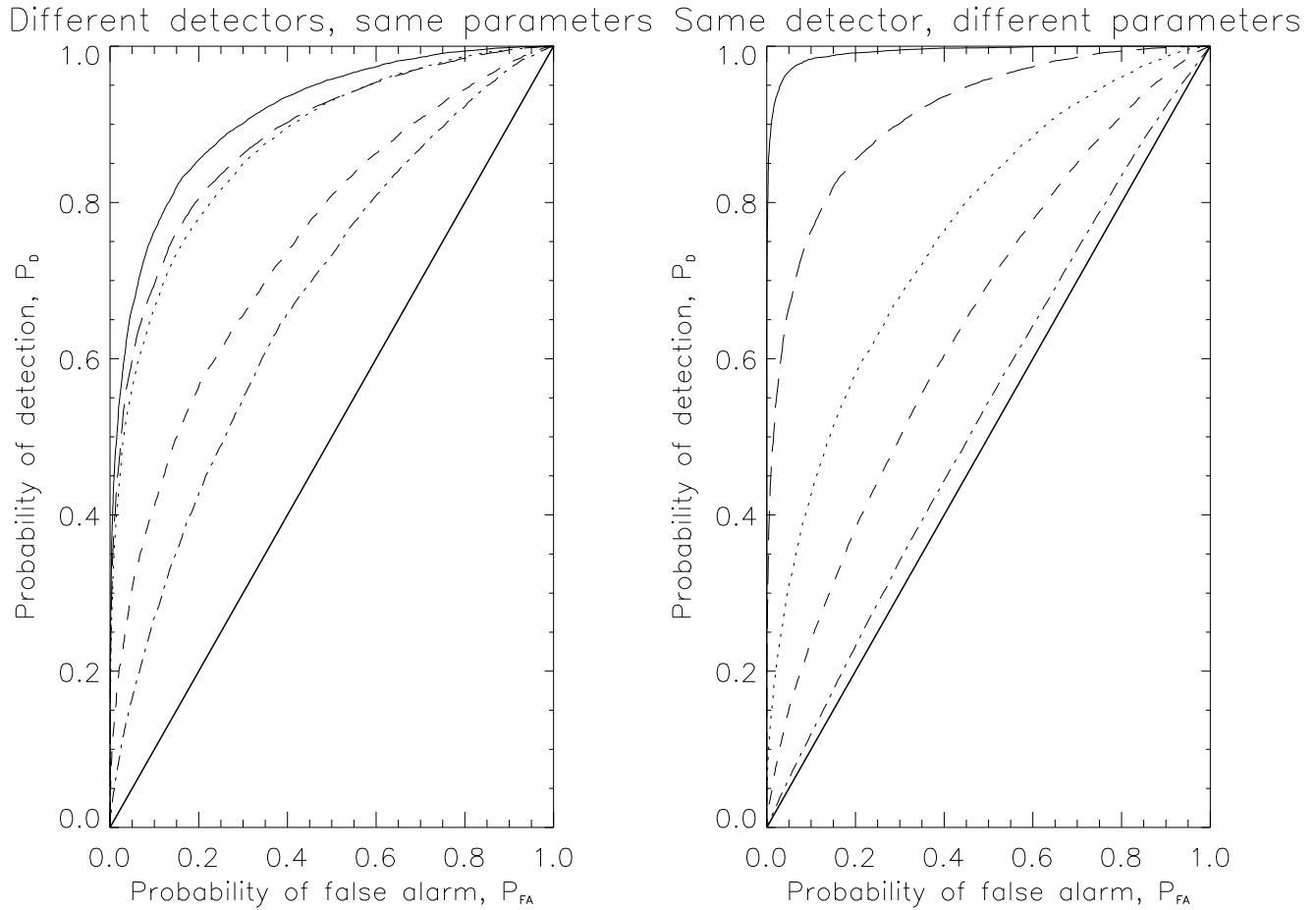}
   \caption{The plot on the left shows the Receiver Operating Curves
for all five detectors used in the simulations for the fixed set of parameters
$(N_{tr} = 3, r_{{\rm obs}} = 5, d = 0.5)$. The curved solid line is the
matched filter, the long dashed line is the correlation, the dotted line is
Deeg's approach, the short dashed line is the BLS and the dot-dashed line
is the Bayesian. The straight solid line shows the limit that corresponds to
random guessing. The plot on the right shows the same diagram for the matched
filter approach and different transit parameters. In the format $(N_{tr},
r_{{\rm obs}},d)$, the curved solid line is $(6,5,0.5)$, the long dashed line
is $(3,5,0.5)$, the dotted line is $(6,5,0.25)$, the short dashed line is
$(3,5,0.25)$ and the dot-dashed line is $(3,5,0.1)$. The lines from the other
sets of input parameters tested would be nearly indiscernibly crowded into
the upper left-hand corner of the diagram, as they represent stronger transit
signals.}
    \end{figure*}

   \begin{figure*}
   \centering
   \includegraphics{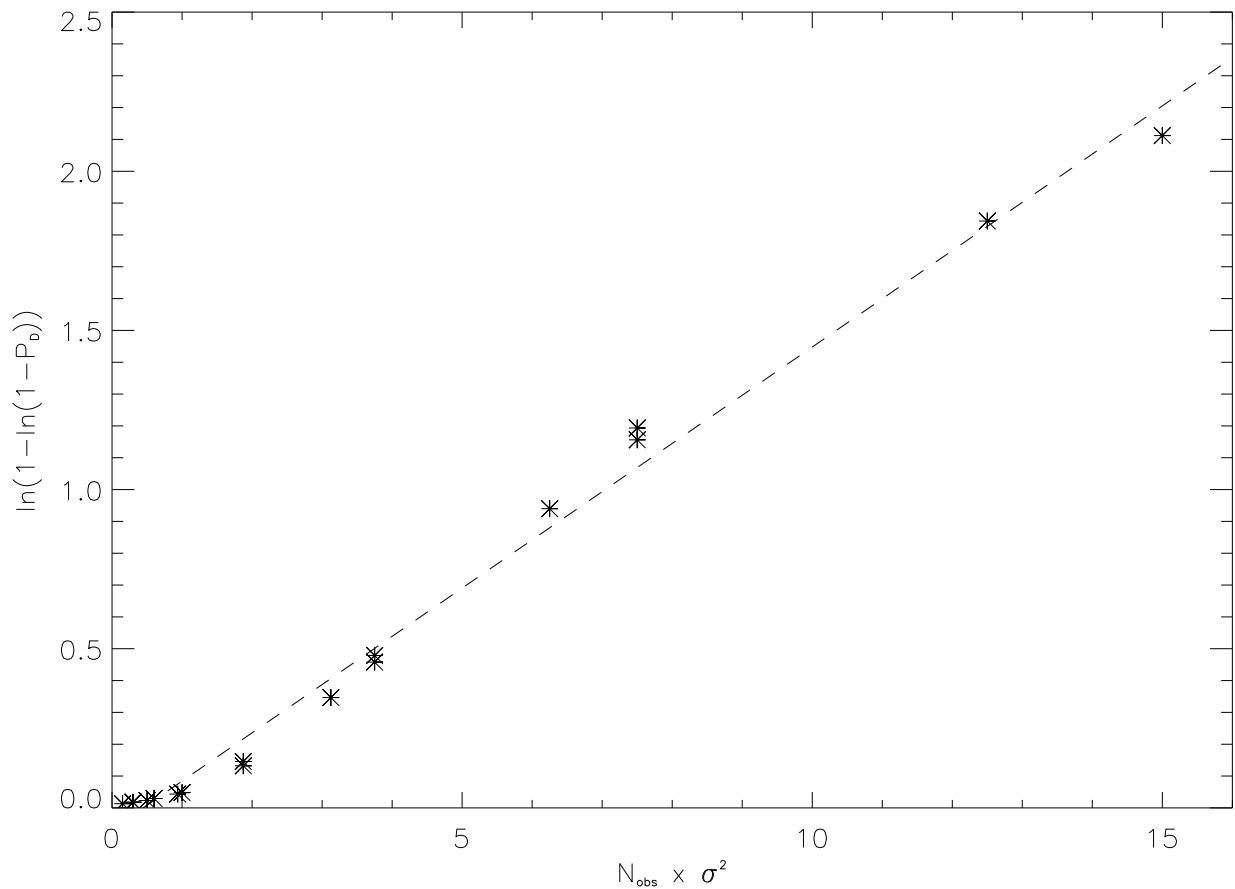}
   \caption{A plot of $\ln(1-(\ln(1-P_{{\rm D}}))$ against $N_{{\rm obs}}
\times \sigma^{2}$, where $P_{{\rm D}}$ is the probability of detection for a
false alarm rate of $10^{-2}$, $N_{{\rm obs}}$ is the number of observations
while the planet is in transit and $\sigma$ is the depth of the transit
observed. The values in this plot are taken from the matched filter approach
column in Table 1. The point corresponding to $(10,10,0.5)$ is missing
because, due to the limited number of events in the simulation, its
$P_{{\rm D}}$ equals one and is therefore undefined in the function used. The
asterisks show the actual values from table 1, while the dashed
line (with a slope of 0.1668) is the best fit straight line to these values.}
    \end{figure*}

\begin{table*}
 \centering
 \caption[]{This table contains the results from verification simulation,
which used the input parameters $(3,3,0.5)$. The first column lists the
detectors. The second column lists the probabilities of detection from
Table 1 for the same input parameters. The third column lists the probability
of detection based on $10^{6}$ iterations with a FAP of 1\%, while the
accompanying errors are the standard deviation calculates from the 100
subsets of $10^{4}$ iterations in the total sample. The last
two columns show the probabilities of detection with false alarm
probabilities of $10^{-3}$ and $10^{-4}$ respectively.}

 \begin{tabular}{c|c|ccc}
  \hline
\multicolumn{1}{c|}{} & \multicolumn{1}{c|}{Probabilities} & \multicolumn{3}{c}{False Alarm Probability}\\
Detector & from Table 1 & $10^{-2} \pm {\rm err}$ & $10^{-3}$ & $10^{-4}$ \\
  \hline
BLS            & 0.1393 & $0.1478 \pm 0.0069$ & 0.0519 & 0.0155 \\ 
Bayesian       & 0.0500 & $0.0505 \pm 0.0034$ & 0.0091 & 0.0016 \\
Correlation    & 0.3920 & $0.3842 \pm 0.0128$ & 0.1786 & 0.0670 \\
Matched filter & 0.4590 & $0.4517 \pm 0.0123$ & 0.2313 & 0.0943 \\
Deeg's        & 0.3617 & $0.3804 \pm 0.0114$ & 0.1783 & 0.0713 \\
 \hline
 \end{tabular}
\end{table*}

When using Monte Carlo simulations, more events always yield more reliable
results. Unfortunately, some of the algorithms - in particular the Bayesian -
are computationally intensive. Despite the use of a dedicated machine
with $8\times1.9$ GHz Athlon processors working in parallel, the simulations
took several weeks to run in total, despite holding several of the
key parameters (period and duration) constant. The FAPs
that can be reliably estimated are restricted by the number of events used
in the Monte Carlo simulations. Reducing the FAP produces a broader
distribution in the results for a fixed number of events, which could blur the
distinction between the detectors.
To evaluate detector performance, it is necessary to use as low a FAP as
possible because the distributions are not identical, meaning that one
detector could be better than another at a relatively high FAP but worse
at a lower one.
A balance between these two needs is required. As $10^{4}$ events were
used for the simulations, a FAP of 1\% is suitable, providing a low enough
FAP for a good comparison while keeping the distributions narrow enough to
evaluate detector performance. In order to ascertain the results of this
analysis, a single set of parameters is simulated with $10^{6}$ events,
allowing both for the widths of the distributions to be estimated and for
lower false alarm probabilities to be examined. This will help to verify
that the relative performance of the different detectors remains more or
less constant.

\section{Results}

The simulations were performed and the results are shown in Table 1, which
contains the input parameters (the number of transits, the rate of
observations, and the depth of the transits) and the probabilities of
detection for the different detectors described above. They
demonstrate clearly that the matched filter is the superior detector.
It is followed unexpectedly by the simple correlation and then Deeg's
approach. The BLS and the Bayesian are significantly worse, despite
the fact that these are the only two detectors that have had papers
in the astronomical literature on them specifically. Table 2 demonstrates
that the results in Table 1 are reliable, as the scatter of the distributions
derived is less than the difference between the detectors and that the
rank of the detectors does not change as the FAP is reduced. A pair of ROCs
is given in Figure 2. The left panel shows the response for different
detectors to a single set of input parameters $(3,5,0.5)$ and the right
panel shows the response of a single detector (the matched filter approach)
to different sets of input parameters.

The response of these different detectors to changes in parameters
can also provide insight into their performance. From inspection, or just
from a basic understanding of statistics, it seems likely that the
significant parameters that govern the detectability of a transit
are the number of observations of the planet during transit
and the S/N of those observations. The number of observations during transit
in these simulations is proportional to the rate of observations and the
number of transits observed. The S/N is related to the depth of the
transit in the simulations performed, as one can equate changing the depth
of the transit to changing the S/N of the observations. These two fundamental
parameters are also related in another way, as the S/N is inversely
proportional to the square root of the exposure time.

In order to determine if there might be some definable relationship between
these fundamental parameters and the probability of detection, an
understanding of the latter must be obtained. If the with-transit
distribution is Gaussian, then the probability of detection will be equal
to the integral of this Gaussian from the point of the chosen false alarm
probability to infinity. This has no simple, analytical solution, but it
is
related to the exponential, which provides a starting point for the empirical
investigation. Figure 3 shows a plot of the number of observations during
transit times the square of the S/N (depth) versus $\ln(1-(\ln(1-P_{{\rm D}}))$,
where $P_{{\rm D}}$ is the probability of detection. While the fit is not
perfect, it does strongly suggest that the probability of detection can be
well-estimated from fundamental parameters. This is particularly true at
moderate to high probabilities, which is the region of interest. Hopefully this
will also mean that an estimation like this is possible for the false alarm
probability, although that analysis is beyond the scope of
the simulations performed here, as it would require several orders of magnitude
more iterations.

\section{Conclusion}

A Monte Carlo simulation using randomly generated light curves drawn from
a set of real data to compare the effectiveness of the different planetary
transit identification algorithms has been performed. Of the methods tested,
which
included the matched filter approach, the linear correlation, the Bayesian
of Defa\"{y} et al. (\cite{defay}), the BLS of Kov\'{a}cs et al.
(\cite{kovacs}) and modified matched filter detector of Doyle et al.
(\cite{doyle}), the matched filter demonstrated the best performance, while
the BLS and the Bayesian both performed poorly by comparison.

An important thing to consider before dismissing the BLS and the
Bayesian is that both of these detectors searched not only for the best
phase but also for the best period, while the other detectors searched only
for the best period in these simulations. This extra free parameter could
certainly affect the results of the Monte Carlo simulations, even though the
period search was strongly limited to a region close to the known period.
Additionally, neither one of these techniques has the depth of the transit
per se as a free parameter. However, the true advantage of this was not seen
in the simulations in this paper, as only the known depth of the transit
was searched for. This means that the approaches that had depth as a free
parameter (the matched filter, the linear correlation, and Deeg's approach)
were not entirely accurately represented, as they would otherwise have
required more statistical tests, which would increase the overall level of
the false alarms, reducing the detection probability at a FAP of 1\%. Perhaps
it is not surprising, then, that these three approaches ranked as the three
best. As with Aigrain's approach, further
analysis covering more parameters might be necessary for a completely accurate
comparison. 

The BLS, as stated in its description in section 2.4, makes use of a $\chi^{2}$
fit. Interestingly, an expectation of the relative performance of a
$\chi^{2}$ fit and the matched filter approach can inferred mathematically.
They are in fact closely related, the $\chi^{2}$ fit being one type of purely
statistical test and the likelihood ratio from which the matched filter is
derived being another. It becomes evident when one expands the basic
formulation for the $\chi^{2}$
\begin{eqnarray}
\chi^{2} & = &\sum_{n=1}^{N}\left(\frac{D_{n}-S_{n}}{\sigma_{n}}\right)^{2} \nonumber \\
  & = & \sum_{n=1}^{N}\left[\left(\frac{D_{n}}{\sigma_{n}}\right)^{2}-2\left(\frac{D_{n}S_{n}}{\sigma_{n}}^{2}\right)+\left(\frac{S_{n}}{\sigma_{n}}\right)^{2}\right]
\end{eqnarray}
and compares it term by term with the matched filter. The last term is strictly
defined and can therefore be neglected, as is done in the derivation
of the matched filter from the likelihood ratio, where the same term appears.
The middle term is actually proportional to the matched filter itself. This
means that the only significant difference that remains between the two
detectors is the first term. This term has no dependence on the
test signal and therefore does not improve the
ability of the detector to discern the signal. On average, it will be
approximately equal to $N$, but there will be scatter around this value. This
scatter is ultimately an extra noise source. Therefore, the best
modification one can make to the $\chi^{2}$ fit is
to remove this extra source of noise, leaving it equivalent to the
matched filter. This is exactly what Kovacs et al. do in the derivation of
their test statistic. This means, in effect, the only mathematical
difference between the matched filter and the BLS is that the BLS has
the depth removed as a free parameter. Further simulations are required to
determine if this actually helps, however, as the computational load of
the BLS and the matched filter are quite similar, even including depth
as a free parameter.

The Bayesian approach exhibits several qualities that could affect its
performance, related to the fact that it performs a Fourier fit to the
shape of the transit. Defa\"{y} et al. (\cite{defay}) state that
this is an improvement over the matched filter approaches offered by Jenkins
et al. (\cite{jenkins0}) and Doyle et al. (\cite{doyle}). However, according
to Kay (\cite{kay}), the use of known information to improve
the chance of making the proper decision between
hypotheses is one of the central ideas of the Bayesian philosophy. As the shape
of any planetary transit can be represented very well by square wells of
different depths and widths, it would seem that this information
could be used to improve the chance of making an identification. Furthermore,
this algorithm limits the number of frequencies used to make the fit to
$m=7$, which the authors claim is sufficient. However, when using a
multi-frequency Fourier fit, it is crucial to have enough frequencies to fit
the ultimate transit shape accurately. Otherwise, signal energy will be lost,
hurting the performance of the detector. 

Lastly, in addition to the comparison of
the detectors, there appears to be something of an empirical relation between
the fundamental parameters (the number of observations during transit and
the S/N of those observations) and the probability of detection at a
fixed false alarm probability. The realization that transit
signal energy is defined by the number of observations during transit times
the square of the S/N of those observations is an important result itself,
one that could be used to expand free parameter space later simulations
without drastically increasing computational load.

\begin{acknowledgements}
Many thanks go to Jon Jenkins at SETI Institute for the great improvement
he coaxed out of this paper with his patience and thorough refereeing.
I would like to thank the Danish Natural Sciences Research Council
for financial support. And finally I would like to thank Hans Kjeldsen
and J\o rgen Christiansen-Dalsgaard for all the support they have provided
me over the past two years.
\end{acknowledgements}

\end{document}